\documentclass[prb,twocolumn,groupedaddress,superscriptaddress]{revtex4-1}

\usepackage{eurosym}
\usepackage{amsfonts}
\usepackage{amsmath}
\usepackage{amssymb}
\usepackage{graphicx}
\usepackage{float}

\begin{document}

\title{Stiffness and coherence length measurements of ultra-thin superconductor, and implications to layered superconductors}

\author{Amit Keren}
\email{keren@physics.technion.ac.il}
\affiliation{ Department of Physics, Technion Israel Institute of Technology, Haifa 32000, Israel}
\author{Nitsan Blau}
\affiliation{ Department of Physics, Technion Israel Institute of Technology, Haifa 32000, Israel}
\author{Nir Gavish}
\affiliation{Faculty of Mathematics, Technion Israel Institute of Technology, Haifa, 32000, Israel }
\author{Oded Kenneth}
\affiliation{ Department of Physics, Technion Israel Institute of Technology, Haifa 32000, Israel}
\author{Yachin Ivry}
\affiliation{Department of Material Science and Engineering, Technion Israel Institute of Technology, Haifa 32000, Israel}
\affiliation{Solid State Institute, Technion—Israel Institute of Technology, Haifa 32000, Israel}
\author{Mohammad Suleiman}
\affiliation{Department of Material Science and Engineering, Technion Israel Institute of Technology, Haifa 32000, Israel}
\affiliation{Solid State Institute, Technion—Israel Institute of Technology, Haifa 32000, Israel}

\date{\today}


\begin{abstract}

Based on the London equation, we use a rotor-free vector potential ${\bf A}$, and current measurements by a SQUID, to determine the superconducting Pearl length $\Lambda$, and coherence length $\xi$, of ultra-thin, ring shaped, MoSi films, as a function of thickness $d$ and temperature $T$. We find that $\xi$ is a function of $d$ with a jump at $\xi \sim d \sim 5$nm. At base temperature the superconducting stiffness, defined by $1/\lambda^2=1/(\Lambda d)$, is an increasing function of $T_c$. Similar behavior, known as the Uemura plot, exist in bulk layered superconductors, but with doping as an implicit parameter. We also provide the critical exponents of $\Lambda (T)$.

\end{abstract}


\maketitle


\section*{\label{sec:Intro} Introduction}

In ultra-thin superconducting (SC) films, SC properties often deviate from their bulk counterparts \cite{Wang1996,Semenov2009,IvryPRB14}. Upon decreasing the SC thickness $d$, there is a suppression of the critical temperature $T_{c}$, the critical magnetic field $H_{c}$, and the critical current density $j_{c}$. The entire SC phase diagram shrinks in all experimental dimensions but does not necessarily disappear \cite{zhang2010superconductivity}. In particular,  bulk superconductor repel small enough magnetic fields from its interior on the penetration depth length scale $\lambda$. The repulsion is due to SC current which flows close to the surface perpendicular to the field and creates an opposite magnetic field to the applied one, leading to a total zero field in the sample's interior. For a 2D SC film, with a field perpendicular to the film ${\bf B}_{\perp}$, super-current can only flow on a quasi 1D edge, therefore,  ${\bf B}_{\perp}$, is expected to easily, but not freely, penetrate into the sample. Magnetic induction parallel to a current carrying 2D (atomically thin) SC sheet, ${\bf B}_{\parallel}$, is not well defined inside the superconductor since it jumps between the two sides of the sheet. 

Nevertheless, surface super-current density $\boldsymbol{\mathcal{J}}$ and the vector potential are well defined even for a 2D superconductor. Therefore, the gauge-invariant London equation is given by
\begin{equation}
 \boldsymbol{\mathcal{J}}=\frac{\psi^2}{\mu_{0}\Lambda\psi_0^2} \left(\frac{\Phi_0}{2\pi} \boldsymbol{\nabla} \phi -{\bf A}_{tot} \right)
\label{eq: 2D London}
\end{equation}
where ${\bf A}$ is the total vector potential, $\Lambda$ is the Pearl length \cite{Pearl1964}, $\Phi_0$ is the SC flux quanta, and $\psi$ and $\phi$ are the magnitude and phase of the SC order parameter respectively. For a SC of thickness $d \ll \lambda$, the Pearl length is given by $1/\Lambda=d/\lambda_0^2= \mu_{0} \psi^2_0 e^{*2} d/ m^{*}$, where $e^*$ and $m^*$ are the carriers charge and mass respectively, and $\psi_0$ is the equilibrium value of the order parameter in the absence of fields.

Equation \ref{eq: 2D London} has a range of validity; \textit{e.g.} $\phi$ cannot change by more than $2\pi$ over inter-atomic distance. Therefore, $\boldsymbol{\mathcal{J}}$ is limited by $\boldsymbol{\mathcal{J}}_c$ set by the distance $\xi$ over which $\phi$ changes by $2\pi$ in a particular material. Interesting questions regarding thin superconductors are how does $\xi$ depend on $d$, and what is the relation between the 2D and 3D stiffness $1/\Lambda$ and $1/\lambda^2$, and $T_c$. These questions have been addressed previously by A.C. methods only \cite{GubinPRB2005}.

To address these questions experimentally we use a Stiffnessometer. Details and drawings of the apparatus can be found in Ref.~\cite{MangelPRB20}. An ideal Stiffnessometer is made of an infinitely long excitation-coil (EC) piercing a ring-shaped SC sample with inner and outer radii $r_{in}$ and $r_{out}$, respectively. Driving a current $I$ through this coil generates a uniform magnetic field in its interior, parallel to its symmetry axis, and zero magnetic field outside. Nevertheless, there is a vector potential ${\bf A}_{ec}$ outside the coil. In the Coulomb gauge ${\bf A}_{ec}=\frac{\Phi_{ec}}{2\pi r}\hat{\theta}$ where $\Phi_{ec}$ is the flux in the EC, and $r$ is the distance from the coil's symmetry axis. According to the London equation, ${\bf A}_{ec}$ leads to a current in the SC ring, which, in turn, generates its own vector potential ${\bf A}_{sc}$. A circular pick-up loop, of radius $R_{pl}$, connected to a SQUID surrounds both the excitation coil and the ring, and measures the total flux through it. This flux can be expressed by ${\bf A}_{tot}={\bf A}_{ec}+{\bf A}_{sc}$ experienced by the pick-up loop. Taking Ampere's law, Eq.~\ref{eq: 2D London}, cylindrical symmetry, and substituting $\psi / \psi_0 \rightarrow \psi$, one gets for $A \equiv  2 \pi R_{pl} A_{sc}/ \Phi_0$, at $r>0$ the equation:
\begin{equation}
A_{rr}+\frac{A_r}{r}-\frac{A}{r^2}+A_{zz}=\frac{\psi^2}{\Lambda}\left(A+\frac{J-m}{r}\right)\delta (z),
\label{eq: Pearl equation}
\end{equation}
with boundary conditions $A_{\rm sc}(0,z)=A_{\rm sc}(\infty,z)=0$. In this equation $r$ and $\Lambda$ are measured in units of $R_{pl}$, $\psi(r)$ is normalized to 1 at it's maximum, $J \equiv \Phi/\Phi_0$ is the normalized excitation coil flux, and the integer $m$ is defined via the relations $\nabla\phi=m\hat\theta/r$.

When cooling a superconductor at zero ${\bf A}$ below $T_c$, $\phi$ is uniform ($m=0$) to minimize the kinetic energy. Since $m$ is quantized, $\phi$ remains uniform upon slightly increasing $J$. This relation holds until $J$ reaches a critical value $J_c$ where $\psi$ is destroyed at least along a path on which flux lines can flow out of the sample, and $\phi$ can change. $\psi$ is controlled by the second Ginzburg-Landau equation
\begin{equation}
-\psi_{rr}-\frac{\psi_{r}}{r}+A^2\psi=\frac{1}{\xi^{2}}\left(\psi-\psi^{3}\right).
\label{eq: 2nd GL}
\end{equation}
Outside of the SC $\psi=0$, and the boundary conditions inside the SC are $\psi_r(r_{in},0)=\psi_r(r_{out},0)=\psi_z(r,\pm d/2)=0$. Therefore, $J_c$, is determined by the coherence length $\xi$. Thus, a measurement of $A_{sc}$ versus $J$ provides information on both $\Lambda$ and $\xi$ simultaneously.

\begin{figure}
	\includegraphics[width=0.45\textwidth]{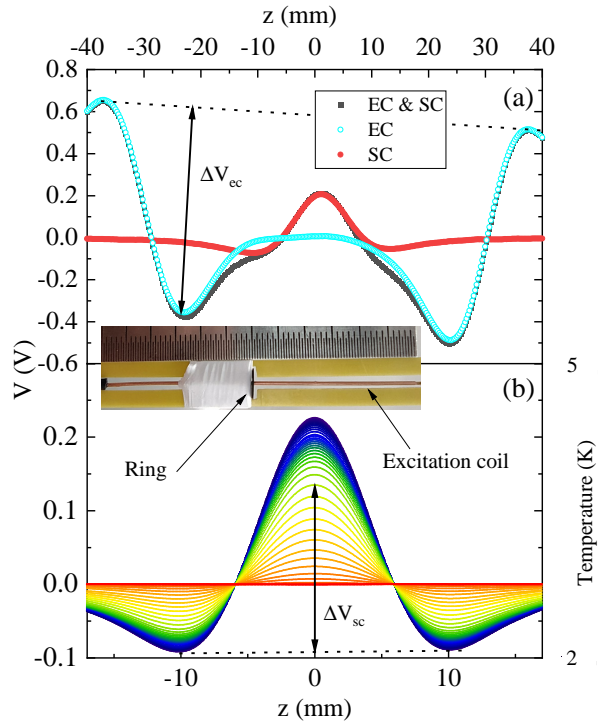}
	\caption{{\bf Experimental configuration} (a) Typical signals obtained with a Cryogenic magnetometer with $80$~mm motion amplitude from: the excitation coil and superconductor below $T_c$ (solid black), the excitation coil alone above $T_c$ (open cyan). The difference is the superconductor contribution alone, (solid red). During the measurements the ring and excitation coil are moving along the $z$ direction in and out of the gradiometer, which is fixed at $z=0$, and the SQUID's voltage is registered. There is a linear drift in the signal during the motion demonstrated by the dotted line. $\Delta V_{ec}$ is the difference between the maximum and the minimum voltage of the EC signal, at specific temperature, as demonstrated by the arrow. (b) The SC signal of $7$~nm thick MoSi film at various temperatures taken with QD MPMS3 magnetometer. $\Delta V_{sc}$ is the difference between the maximum and the minimum voltage of the SC signal as demonstrated by the arrow. Upon increasing the temperature from $T=2$~K, the SC signal decreased until it vanishes above $T_{c}\approx 5$~K. The inset is a picture of a ring and the excitation coil. A full description of the apparatus is given in Ref.~\cite{MangelPRB20} }
	\label{fig: Experiment}
\end{figure}

\section*{\label{sec:Exp} Experiment}

We have implemented the Stiffnessometer in a QD-MPMS3 magnetometer and its $^3$He insert. For the standard operation, the EC is $60$~mm long, copper, double-layered coil with an external diameter of $0.8$~mm and a total of $1940$ windings. In the $^3$He system we used a 30~mm long double-layerd superconducting NbTi coil, with an external diameter of 0.2~mm, and a total of 1200 windings. The SC MoSi film is grown on Si-ring substrate with an oxide layer of 175~$\mu$m, $r_{in}=0.5$~mm and $r_{out}=2.5$~mm and is placed around the center of the coil (see picture in Fig.~\ref{fig: Experiment}). The pickup loop is, in fact, a second-order gradiometer. It is made of three winding groups. The two outer groups are wounded clockwise, and the inner group is wounded anticlockwise. The radius of the gradiometer is $R_{pl}=8.5$~mm. In the measurement, the gradiometer is static and its center is fixed at $z=0$, while the coil and the SC ring move rigidly in the ${\bf \hat{z}}$ direction. Therefore, the magnetic flux through the gradiometer is a function of $z$. In QD-MPMS3 the motion is limited to $\pm30$~mm. The ultra low field option of the MPMS3 is used to minimize the external field down to 0.02~Oe.

The finite coil and gradiometer's geometry yields a unique SQUID voltage output as demonstrated in Fig.~\ref{fig: Experiment}(a). This particular measurement was taken with a Cryogenic SQUID where a motion of $\pm 40$~mm is possible, allowing to capture all important features of the signal symmetrically around $z=0$ in one scan. The EC contribution to the signal is obtained from the measurement above $T_{c}$ and is demonstrated by the cyan open symbols in the figure. The total signal below $T_c$ is given by the black solid symbols. By subtracting the EC signal from the total signal, we obtain the SC signal in red solid symbols in panel (a) and as a stand alone in panel (b). The difference between the maximum and minimum voltage of the SC signal is defined as $\Delta V_{sc}$. Similarly, $\Delta V_{ec}$ is defined for the EC signal. Both are depicted in the figure. The measurable voltages $\Delta V_{sc}$ and $\Delta V_{ec}$ are proportional to the flux through the gradiometer generated by the superconductor and EC respectively.

\begin{figure}
	\includegraphics[width=0.5\textwidth]{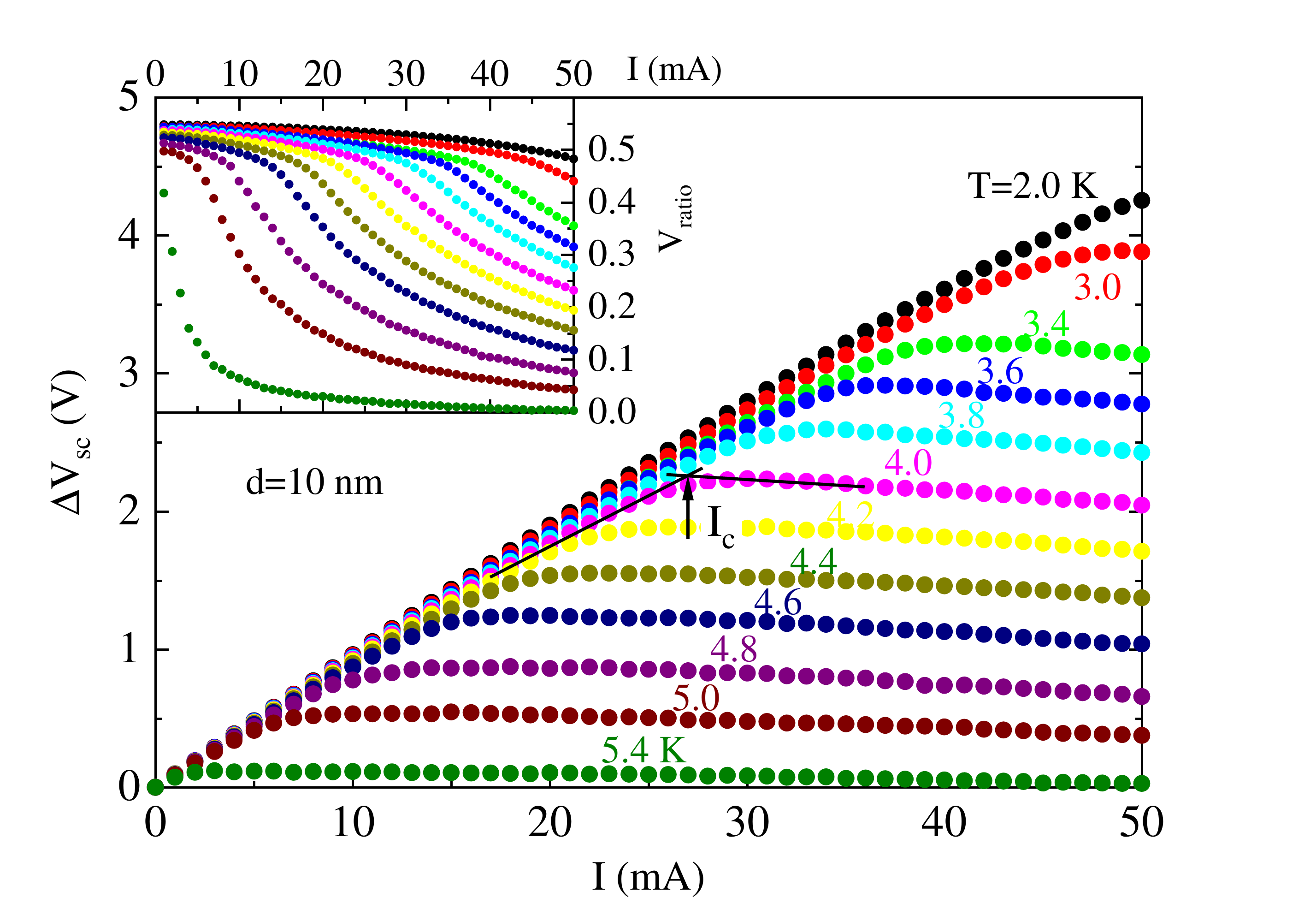}
	\caption{ \textbf{Critical Current} measurements of $10$~nm thick MoSi film at different temperatures showing $\Delta V_{sc}$ as a function of the current $I$ in the excitation coil. The critical current is determined by the break point demonstrated on the $T=4$~K data. The inset is the normalized signal $V_{ratio}=\Delta V_{sc}/\Delta V_{ec}$}
	\label{fig: Critical-Current}
\end{figure}

\section*{\label{sec:Res} Results}

Fig.~\ref{fig: Experiment}(b) presents the SC signal of a $7$~nm thick MoSi film as a function of temperature. The signal resembles that of a point-like magnetic moment, and diminishes smoothly upon heating towards the film's $T_c$. From such a measurement we extract $\Delta V_{sc}$ at each temperature. The error in $\Delta V$ are estimated from the difference between the left and right minimum of the signal. It is enough to determine $\Delta V_{ec}$ once per current.

Fig.~\ref{fig: Critical-Current} presents $\Delta V_{sc}$ measurements as a function of the coil current $I$ at various temperatures. The inset shows $V_{ratio} \equiv \Delta V_{sc}/\Delta V_{ec}$ as a function of $I$. The critical current $I_{c}$ is defined as the break point between the ring's vector potential (proportional to $\Delta V_{sc}$) and the EC current $I$. This point is demonstrated in the figure on the $T=4$~K data. The cooler the sample is, the bigger is the break point current, namely, the critical current increases as the temperature decreases, as expected.

\begin{figure}
	\includegraphics[width=0.5\textwidth]{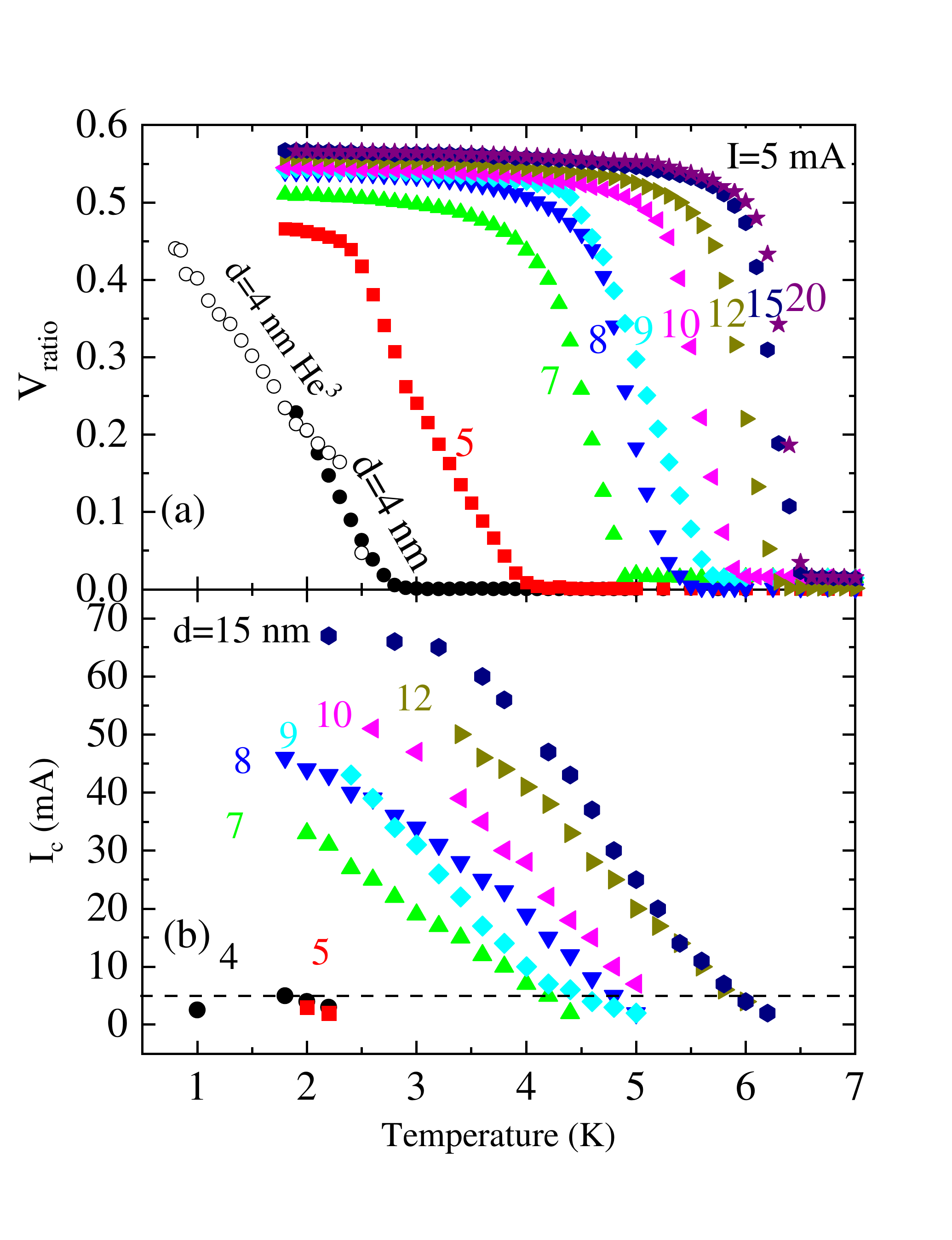}
	\caption{\textbf{Raw data} from MoSi rings as a function of temperature and various thicknesses $d$ (a) at an applied excitation coil current of $I=5$~mA, and (b) of the critical currents. The open symbol in (a) are He$^3$ measurements}
	\label{fig: Vratio and Ic vs T} 
\end{figure}

Fig.~\ref{fig: Vratio and Ic vs T}(a) presents $V_{ratio}$ measurements of MoSi films prepared with different thicknesses $d$, as function of the temperature and applied EC current of $I=5$~mA. As can be seen, $T_{c}$ increases with increasing thickness and saturates for $d>15$~nm. Similarly, $V_{ratio}$ increases with increasing thickness and saturates. For the sample with $d=4$~nm, $V_{ratio}$ did not saturate at the temperatures achievable with a standard MPMS3 cryostat and the experiment was extended using the $^3$He refrigerator. The movement of the $^3$He cryostat relative to the gradiometer is limited and $\Delta V_{ec}$ cannot be determined. Therefore, we scaled the  $^3$He measurements to match the high temperature ones. Due to the extra coil, leads, and current, the $^3$He did not cool below $1$~K, and even at this temperature there is no clear saturation of the signal.

In Fig.~\ref{fig: Vratio and Ic vs T}(b) we show the critical current $I_c$ in the EC, obtained from Fig.~\ref{fig: Critical-Current} type measurements, as a function of temperature and different $d$ values. Naturally, $I_c$ decreases with decreasing thickness and increasing temperature. It seems to collapse, though not to zero, when $d$ is lower than $7$~nm. Since the $5$~mA current at which panel (a) data is taken, is slightly above the critical current of the $4$ and $5$~nm films, the reported $V_{ratio}$ for these samples in panel (a) is a slight underestimate of $V_{ratio}$ for current below $I_c$.

\begin{figure}
	\includegraphics[width=\columnwidth]{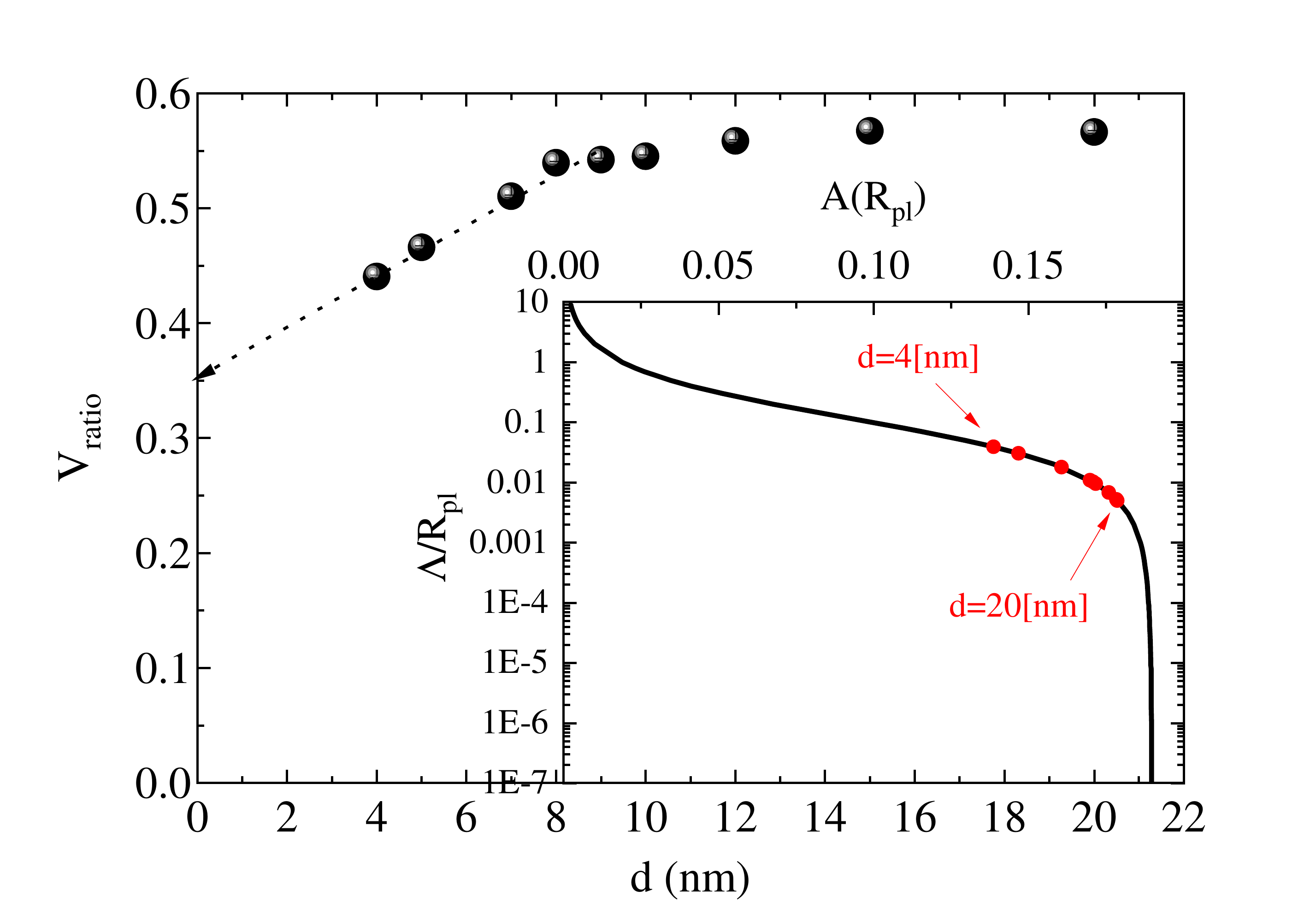}
	\caption{\textbf{The signal intensity} $V_{ratio}$, at the lowest temperature, as a function of the film thickness $d$. Linear extrapolation demonstrated by the arrow suggest that the signal should be detectable even at the level of a single atomic layer of any material with similar 3D stiffness. The inset shows the numerical solution of Eq.~\ref{eq: numerical}, evaluated at $R_{pl}=8.5$~mm, for different Pearl lengths $\Lambda$, and a ring with $r_{in}=0.5$~mm and $r_{out}=2.5$~mm. The conversion of the lowest temperature data from $A$ to $\Lambda$, for each film, is demonstrated by red circles. }
	\label{fig: Vratio vs d and PDE}
\end{figure}

It is interesting to follow $V_{ratio}$ at the lowest temperature as a function of $d$ since this might predict what would the signal be had we have been able to grow a single unit cell. This is shown in Fig.~\ref{fig: Vratio vs d and PDE}. The arrow is an extrapolation to $d$ of a single unit cell. The extrapolation suggests that even in a single unit cell sample the signal should be detectable. 


\section*{\label{sec:Anal} Analysis}

\begin{figure}
	\includegraphics[width=0.5\textwidth]{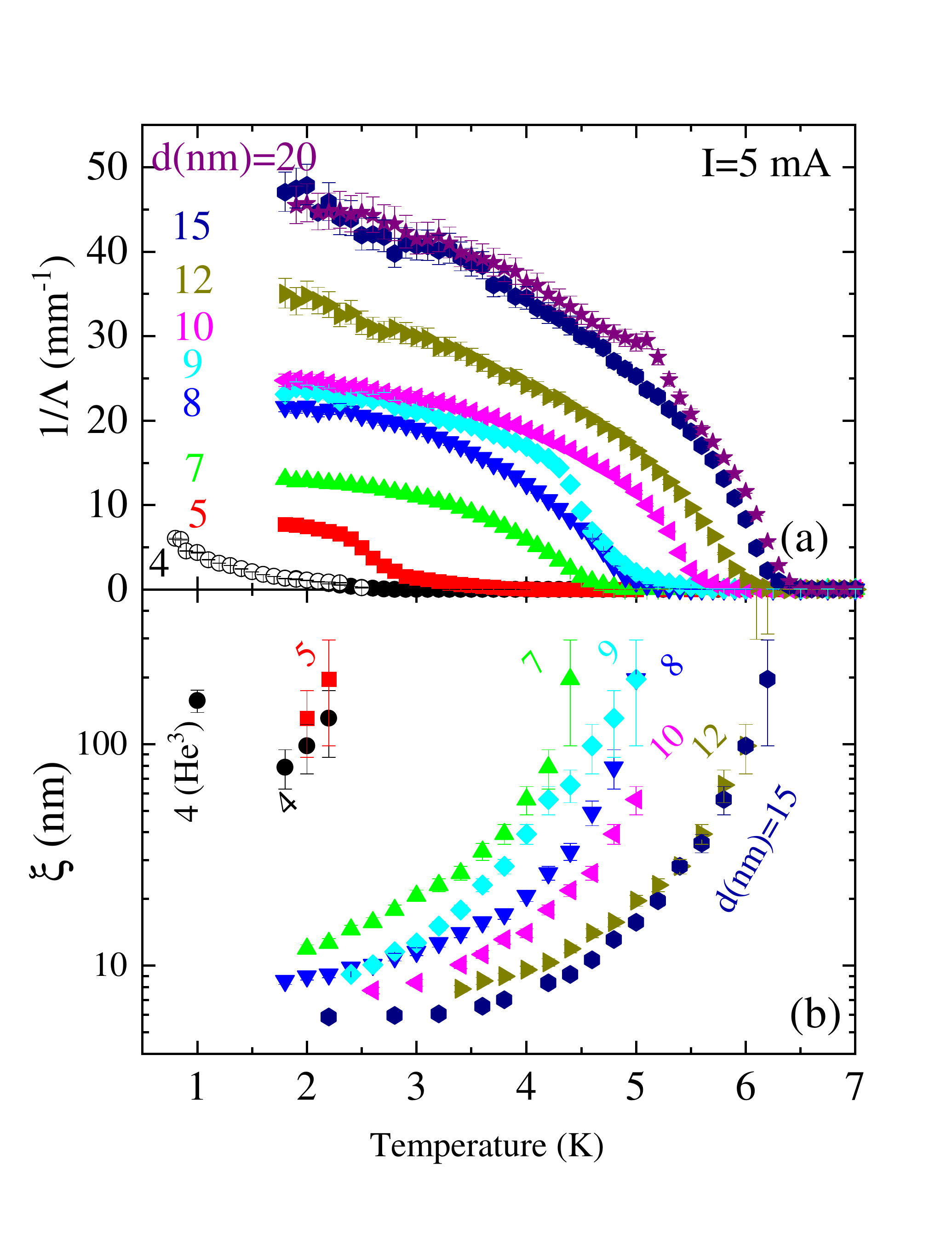}
	\caption{ \textbf{Superconducting lengths} (a) Pearl $\Lambda$ and (b) coherence $\xi$, as a function of temperature and ring thickness. All data sets are measured with a current of $5$~mA at the excitation coil. The data taken with a $^3$He insert are presented by the open symbols.}
	\label{fig: Pearl and xi vs T}
\end{figure}

The SQUID is measuring flux through the gradiometer, which for a point like sample is determined by the magnetic dipole moment.  Since this flux is proportional to the vector potential produced by the element passing through the gradiometer, the measured voltages $\Delta V_{sc}$ and $\Delta V_{ec}$ are proportional to $A_{sc}(R_{pl},0)$ and $A_{ec}(R_{pl},0)$ respectively. Therefore, 
\begin{equation}
V_{ratio}=GA(R_{pl},0).
\label{eq: V to A conversion}
\end{equation}
The proportionality constant $G$ can be found numerically based on the gradiometer parameters \cite{MangelPRB20}, or experimentally as we explain shortly. For the MPMS3 gradiometer, we found numerically that $G\approx 2.9$. The experimental value is reported below.

The analysis of our data is done by solving Eqs. \ref{eq: Pearl equation} and \ref{eq: 2nd GL}, for the Stiffnessometer setup, in two relatively simple limits: (I) $J\to 0$, namely currents in the superconductor are weak and $\psi=1$ everywhere, and (II) $J\to J_c$ meaning $\psi=0$ in most of the superconductor. Experimentally, limit (I) means $I \ll I_c$ and limit (II) is $I \rightarrow I_c$.

In practice, in limit (I), we solve numerically for a range of $\Lambda$'s Laplace's equation $ \nabla^{2}{\bf A}=0 $ in a box $0<z,r<L$, where $L$ is the box length. The ring is located at the box edge $z=0,\;r_{in}<r<r_{out}$, hence the boundary conditions (obtained from the magnetic field jumps) are:
\begin{equation}
\begin{cases}
\frac{\partial A}{\partial z}=\frac{1}{2\Lambda}\left(\frac{1}{r}+A\right) & z=0\;,\;r_{in}<r<r_{out}\\
\frac{\partial A}{\partial z}=0 & z=0\;,\;0<r<r_{in}\\
\frac{\partial A}{\partial z}=0 & z=0\;,\;r_{out}<r<L.\\
\end{cases}
\label{eq: numerical}
\end{equation}
$J$ was absorbed into the normalization of $A$. We solve these equations using finite-elements method with $\mathrm{FreeFem++}$ and $\mathrm{Comsol-5.3a}$ softwares for comparison. 

From the numerical solutions we obtain $A(R_{pl},0)$ for different values of $\Lambda$ as shown in the inset of Fig.~\ref{fig: Vratio vs d and PDE}. There are two saturation regions ($\Lambda<0.001$ and $\Lambda>10$). The short Pearl length saturation region corresponds to the thick samples at low temperatures where $V_{ratio}$ is no longer temperature dependence (see Fig.~\ref{fig: Vratio vs d and PDE}). By comparing $V_{ratio}$ of a $d=150$~nm Nb sample at $T\rightarrow 0$ to the numerical maximum magnitude of $A(R_{pl},0)$, we determine $G=3.18$ experimentally, which is similar to the calculated value. Note that when the Stiffnessometer is saturated it loses its sensitivity to $\Lambda$.

\begin{figure}
	\includegraphics[width=0.5\textwidth]{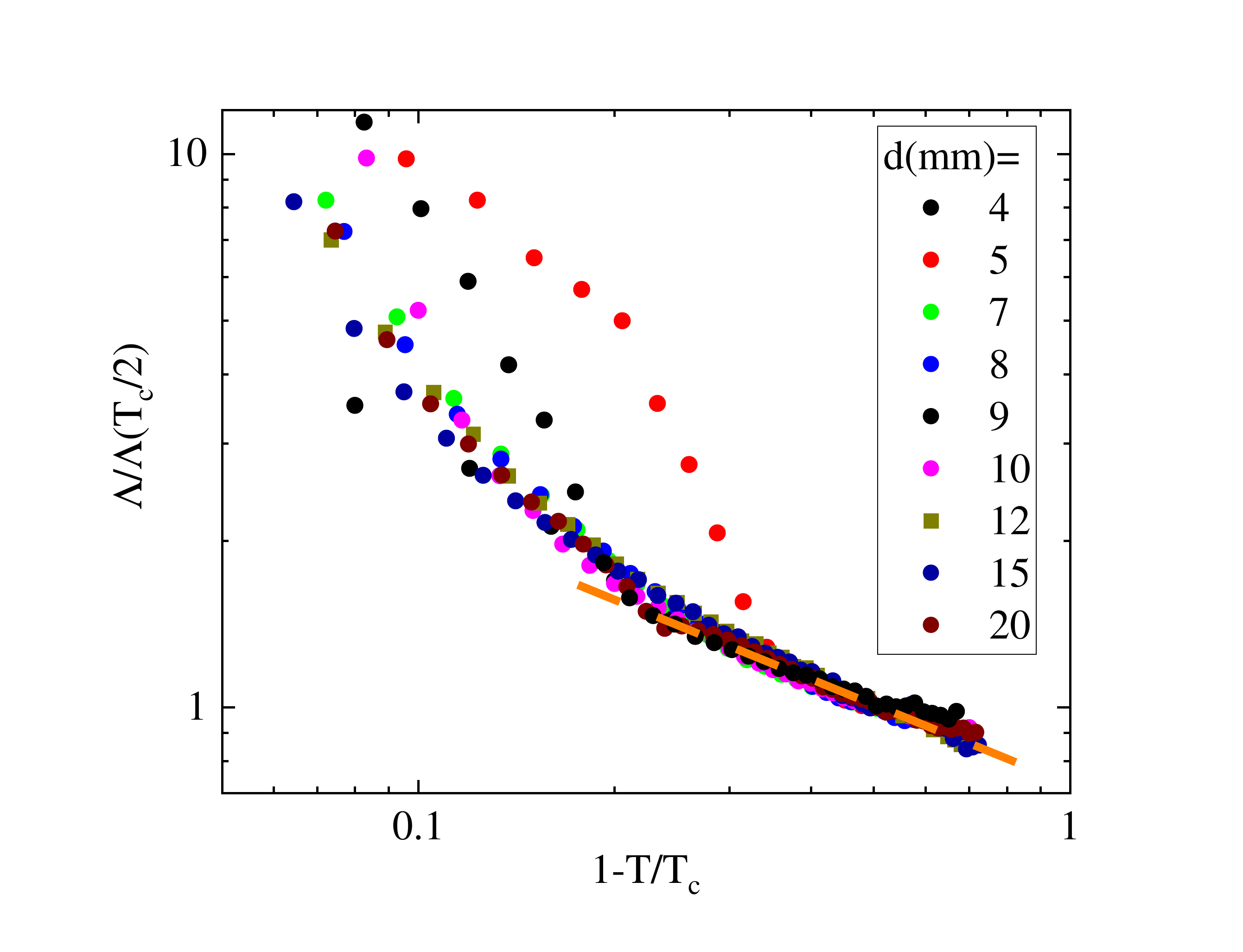}
	\caption{ \textbf{Scaling} A log-log plot of $\Lambda$ normalized at $T_c/2$, as a function of $1-T/T_c$ for thicknesses marked in the legend. A single power law does not fit the entire data set. In the region where the determination of $\Lambda$ is done with a current lower than $j_c$, marked by the orange dashed line, the power is $\gamma=1.50 \pm 0.05 $}
	\label{fig: Scaling}
\end{figure}

Using the measured $V_{ratio}$ in Fig.~\ref{fig: Vratio and Ic vs T}(a), Eq.~\ref{eq: V to A conversion}, and the inset of Fig.~\ref{fig: Vratio vs d and PDE} we extract the Pearl length, measured with $I=5$~mA, as a function of the temperature and sample thickness, and depict its inverse $\Lambda\left(T\right)^{-1}$ in Fig.~\ref{fig: Pearl and xi vs T}(a). This procedure is demonstrated, for the lowest temperature of each sample, by the red circles in Fig.~\ref{fig: Vratio vs d and PDE} inset. The 2D stiffness shows second order phase transition, and increases with increasing thickness. The determination of $T_c$ becomes increasingly inaccurate as $d$ decreases. $\Lambda(T)$ normalized by $\Lambda(T_c/2)$ as a function of $1-T/T_c$ is presented in  Fig.~\ref{fig: Scaling}. At temperatures far enough from $T_c$, the applied current is lower than the critical current and the determination of $\Lambda$ using limit (I) is accurate. This region is marked by the orange dashed line. In this region, all data sets can be made to collapse into a single function by small adjustments of $T_c$ within the uncertainty of its determination from Fig.~\ref{fig: Pearl and xi vs T}(a). A power law of the form $\Lambda/\Lambda(T_c/2)=E+F(1-T/T_c)^{-\gamma} $, with $\gamma=1.5\pm 0.05$ fits the data well.

\begin{figure}
	\includegraphics[width=0.5\textwidth]{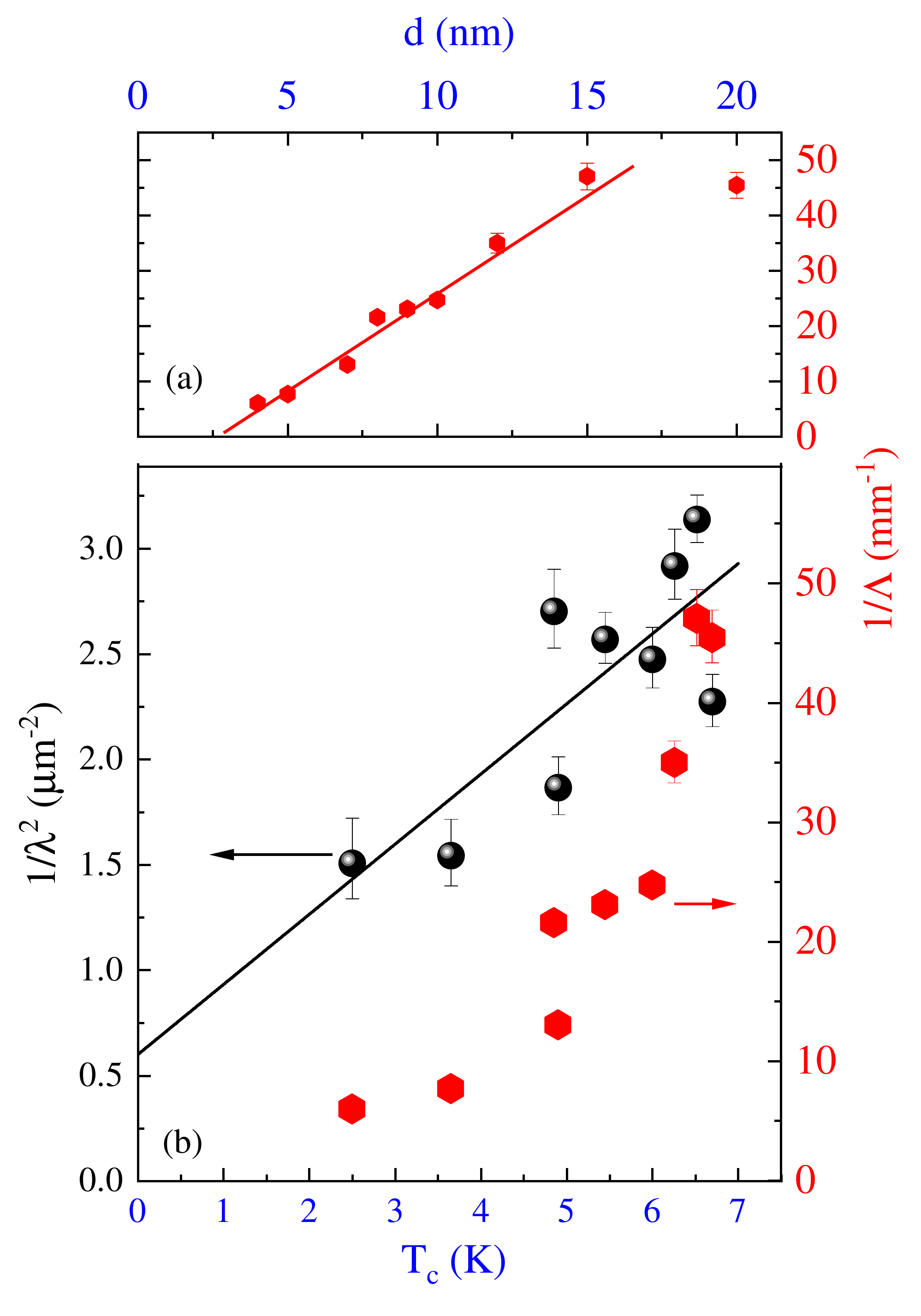}
	\caption{ \textbf{Base temperature 3D and 2D stiffness $\lambda^{-2}$ and $\Lambda^{-1}$} , respectively plotted vs. film thickness $d$ (a) and $T_c$ (b). The slopes of the fitted solid lines are: (a) $3.5\pm 0.2$~$\mu$m$^{-2}$ and (b) $0.3\pm 0.1$~$\mu$m$^{-2}$K$^{-1}$. The error bars on $\lambda^{-2}$ include a $0.5$~nm error in $d$.}
	\label{fig: Summary}
\end{figure}

Data analysis in limit (II) relies on the current density in the SC being strongest in the inner radius of the ring \cite{Gavish2020}. Therefore, destruction of the order parameter starts there and propagate to the outer radius as $J$ increases. We assume that cylindrical symmetry is respected and no vortices enter the sample since no external field is applied, and since $\Delta V_{sc}$ is proportional to the current nearly up to $I_c$ as it should be without vortices. This assumption requires experimental verification. Nevertheless, under these assumptions we expect the critical current $I_c$ to show up when $\psi  \to 0$ in the entire SC. This allows one to linearize Eq. \ref{eq: 2nd GL} and to approximate ${\bf A}$ by the coil's vector potential. In this case,
\begin{equation}
	-\psi_{rr}-\frac{\psi_{r}}{r}+A_{ec}^2\psi=\frac{1}{\xi^{2}} \psi .
	\label{eq: linear 2nd GL}
\end{equation}
The existence of solution which decays rather than blow up at small $r$ requires $A_{ec} \leq 1/\xi$. This leads to a critical flux
\begin{equation}
	J_{c}\equiv\frac{\Phi_{c}}{\Phi_{0}} \simeq \frac{r_{out}}{\xi}
	\label{eq: Jc}
\end{equation}
where $\Phi_{c}=\mu_{0}n\pi R_{ec}^{2}I_{c}$, $n$ is the coil's windings density, and $R_{ec}$ is its radius. Eq. \ref{eq: Jc} and its corrections are discussed in detail in App.~\ref{sec: App}. A second analysis strategy in limit (II) is to estimate at which current does the linearity between the signal and applied current breaks. An estimate of this current, for a narrow ring, gives a factor of $\sqrt{3}$ correction in Eq.~\ref{eq: Jc}, which is not noticeable on a log scale. A full analysis of limit (II) will be given elsewhere. Using Eq.~\ref{eq: Jc} we extract $\xi$ as shown in Fig.~\ref{fig: Pearl and xi vs T}(b). It is clear that $\xi$ increases when $T$ increases, and as $d$ decreases. There seems to be a jump in $\xi$ at $d$ which obeys $\xi(d) \simeq d$. Unfortunately, the data is not systematic enough between samples to fit them with a single power law. 

Comparing our results with different measurements of thin films such as $\delta-\mathrm{NbN}$ \cite{Kamlapure2010} or MoGe \cite{draskovicPRB2013,mandalPRB2020}, we find a similar film-thickness dependency of $T_{c}$ and penetration depth $\lambda$ order of magnitude. However, while in our and Ref.~\cite{draskovicPRB2013} and ~\cite{mandalPRB2020} measurements the stiffness monotonically increases with decreasing temperature, in $\delta-\mathrm{NbN}$ Ref.~\cite{Kamlapure2010} it saturates. We speculate that this is not a material issue but rather a fundamental difference between experimental techniques. We also find that our value of $\xi$ in MoSi is on the same order of magnitude as determined from the upper critical field $H_{c2}$ in MoSi \cite{BaoIEEE2021} and MoGe \cite{draskovicPRB2013,mandalPRB2020}. It should be pointed out that in $H_{c2}$ measurements there is some arbitrariness since it is determined as the field at which the resistivity drops by 50\%. Under these circumstances we can only compare order of magnitudes. Moreover, for MoGe the coherence length is thickness independent in Ref. \cite{draskovicPRB2013}. In Ref. \cite{mandalPRB2020} $\xi$ decreases with decreasing thickness only for $d < 5$~nm, and is ascribed to a transition between Fermionic and Bosinic SC. The difference between different $\xi$ measurements could also be due to the difference between global and local superconductivity as detected by the different techniques.

\section*{ Conclusions }

In Fig.~\ref{fig: Summary}(a) we show $1/\Lambda$ as a function of $d$, and in \ref{fig: Summary}(b) 
we depict  $1/\Lambda$ and $1/\lambda^2\equiv 1/(\Lambda d)$ as a function of $T_c$, with $d$ as an implicit parameter. Disregarding a small offset, $1/\Lambda$ is proportional to $d$, over some range of $d$. The offset might have to do with film percolation since we did not manage to obtain a signal from films thinner than $4$~nm. The deviation from proportionality occurs at a thickness scale on the order of $\xi$. Fig.~\ref{fig: Summary}(b) indicates that $1/\Lambda$ is a function of $T_c$ with a power similar to 2. The best description of the 3D stiffness, despite the noise, is with a linear dependence of $1/\lambda^2 $ on $T_c$. A linearity between the 3D stiffness and $T_c$ is found in many bulk layered superconductors \cite{UemuraPRB91} and films \cite{draskovicPRB2015}, with doping as the implicit parameter, and is known as the Uemura plot. Our result suggests that doping might affect the effective dimensionality of layered superconductors, via the formation of overlapping SC island in different layers, leading to the observed relations between the 3D stiffness and $T_c$. The possibility that granularity plays an important role in the Uemura plot was suggested previously \cite{ImryPRL2012}.  

We also found in this work that $\xi$ is a function of $d$ with a jump when $d \sim \xi$, and that a Stiffnessometer might be sensitive enough to measure the Pearl and coherence lengths of single atomic layer superconductor.


\section*{acknowledgments}
This research is supported by the Israeli Science Foundation (ISF) personal grant No. 315/17, ISF MAFAT quantum science and technology grant 1251/19, by the Russel Berrie Nanotechnology institute, Technion, and by the Nancy and Stephen Grand Technion Energy Program. YI acknowledges financial support from the ISF grants number 1602/17 and the Zuckerman STEM Leadership Program. We are grateful to Guy Ankonina from the Technion Photovoltaic Laboratory, RBNI-GTEP for the sample preparation.


\section*{ AUTHOR INFORMATION}
A.K. conceived the project and wrote the paper. N.B. constructed the measuring device, performed the measurements, and analyzed the data. N.G. and O.K. developed the theory of App.~\ref{sec: App}. Y.I. and M.S. developed the MoSi preparation method in the Technion.


\appendix

\section{ \label{sec: App} The critical flux}

At very high values of current $I$ (or applied flux) $J$, one expects to have only the trivial solution to the set of Eqs.~\ref{eq: Pearl equation} and \ref{eq: 2nd GL} where $\psi=0$ (and $A=A_{ec}$) everywhere. As $J$ is gradually decreased we expect that there should exist a critical value $J_0$ such that at $J<J_0$ the trivial solution ceases to be an energetically stable solution. Note that it may happen (and our results so far for hollow cylinder support this possibility) that the trivial solution ceases to be the minimal free energy solution already at some higher value $J_1$ of $J$ \cite{Gavish2020}. This would simply mean that in the range $J_0<J<J_1$ there exist an energy barrier between the trivial and the minimal free energy solutions.

A standard way to check whether a given solution is energetically stable is to expand the free energy
functional to quadratic order around the given solution, find its eigenmodes, and verify whether the
corresponding eigenvalues are all positive or not.
In the case of the trivial solution $\psi\equiv0$ the quadratic expansion is particularly simple.
\begin{equation}
	\label{Equad}
	F \simeq \frac{1}{2}\int_{-d/2}^{d/2}\int_{r_{in}}^{r_{out}}\left(\xi^2\left((\nabla\psi)^2+J^2\frac{\psi^2}{r^2}\right)-\psi^2\right)r dr dz
\end{equation}
Putting $\psi(r,z)=\tilde{\psi}(r)\cos(\frac{n\pi}{d}(z+d/2))$ and minimizing leads to $n=0$. Therefore, an eigenmode with minimal eigenvalue of this expression must satisfy $\psi_z \equiv 0$ everywhere. This is consistent with the boundary conditions at $z=\pm d/2$. The $z$-independence imply in particular that the shape of the minimal eigenmode does not
depend on the width $d$.

The equation for an eigenmode with eigenvalue $c$ of the functional (\ref{Equad}) is a Bessel equation
\begin{equation}\label{eig}
	-\xi^2\psi''(r)-\frac{\xi^2}{r}\psi'(r)+\left(\frac{J^2\xi^2}{r^2}-1\right)\psi(r)=c\psi(r)
\end{equation}
We shall continue under the assumption that $\xi/r \ll 1$. If we think of the functional in Eq (\ref{Equad}) as corresponding to a 1D radial Schrodinger equation, then the expression $(\xi^2J^2/r^2-1)$ which multiplies $\psi(r)^2$ 
can be interpreted as the corresponding potential energy.
Appearance of a negative eigenvalue requires this potential energy to become negative at least somewhere.
It therefore follows that we must have $J_0< r_{out}/\xi$.

If $J_0< r_{in}/\xi$ the potential is negative everywhere and substituting $\psi(r)\equiv 1$ gives $F<0$. Therefore, $J_0> r_{in}/\xi$ and hence that $J_0=O(\xi^{-1})$. In the case $\xi\ll r$ we expect however  $J_0$ to be dominated by $r_{out}$ and being practically independent of $r_{in}$ because upon reducing $J$ the SC forms first at the outer rim of the ring \cite{Gavish2020}. This suggests that $J_0=r_t/\xi$ for some $r_{in}<r_t<r_{out}$ which is much closer to $r_{out}$
than to $r_{in}$ (i.e. $r_{out}-r_t\ll r_{out}-r_{in}$). The potential energy then changes sign precisely at $r=r_t$ being negative only over $[r_t,r_{out}]$. The corresponding eigenmode must therefore be decaying fast outside this interval.
One can therefore approximate the potential in the region of relevance by the expression $\frac{2(r_t-r)}{r_t}$.
The equation for $\psi$ then reduces to a Stokes type of equation which is solved by an Airy function.
$$\xi^2(\psi''(r)+\frac{1}{r_{out}}\psi'(r))+2\frac{r-r_t}{r_{out}}\psi(r)=0$$
$$\psi(r)=e^{-\frac{r}{2r_{out}}} Ai\left( \frac{\xi^2-8r_{out}(r-r_t)}{2^{8/3}r_{out}^{4/3}\xi^{2/3}}\right)$$
where we chose the solution which is well behaved (in fact decaying) at small $r$.

The value of $r_t$ (and hence of $J_0$) can now be determined by enforcing the boundary condition $\psi'(r_{out})=0$.
$$r_t=r_{out}\left(1-0.808617{\left(\frac{\xi}{r_{out}}\right)}^{2/3}+O\left({\left(\frac{\xi}{r_{out}}\right)}^{4/3}\right)\right)$$
$$J_0=\frac{r_{out}}{\xi}  -0.808617\left(\frac{r_{out}}{\xi}\right)^{1/3}+O\left(\left(\frac{\xi}{r_{out}}\right)^{1/3}\right)$$
The constant $x_0=-0.808617$ appearing here is the numerical solution of the equation $Ai'(-2^{1/3}x_0)=0$. For typical values of $r$ and $\xi$ we recover Eq.~\ref{eq: Jc} with $J_0=J_c$.


%

\end{document}